# Integrated physics-informed learning and resonance process signature for the prediction of fatigue crack growth for laser-fused alloys

Panayiotis Kousoulas[a,b], Rahul Sharma[a,b], Y.B. Guo[a,b,*]

[a] Dept. of Mechanical and Aerospace Engineering, Rutgers University-New Brunswick, Piscataway, NJ 08854, USA
[b] New Jersey Advanced Manufacturing Institute, Rutgers University-New Brunswick, Piscataway, NJ 08854, USA

**Abstract**
Fatigue behaviors of metal components by laser fusion suffer from scattering due to random geometrical defects (e.g., porosity, lack of fusion). Monitoring fatigue crack initiation and growth is critical, especially for laser-fused components with significant inherent fatigue scattering. Conventional statistics-based curve-fitting fatigue models have difficulty incorporating significant scattering in their fatigue life due to the random geometrical defects. A scattering-informed predictive method is needed for laser-fused materials' crack size and growth. Current data-driven machine learning could circumvent the issue of deterministic modeling, but results in a black-box function that lacks interpretability. To address these challenges, this study explores a novel nondimensionalized physics-informed machine learning (PIML) model to predict fatigue crack growth of laser-fused SS-316L by integrating fatigue laws and constraints with small data to ensure a realistic and interpretable prediction. Resonance process signature data were leveraged with Paris's law to train the PIML model without experimental crack growth data. The results show that Paris's law constants can be learned with good similarity to comparable data from the literature, and the crack growth rate can be predicted to compute crack sizes.

## 1. Introduction

While the advantages of metal additive manufacturing (AM), such as laser powder bed fusion (LPBF) have been well recognized [1], little to no publicity about metal AM being used for critical, load-bearing components has been distributed yet. This is because metal AM processes, especially LPBF, have inherent random geometrical defects (e.g., gas pores, lack of fusion porosity, keyhole pores) and rough surfaces. The former is due to the complex dynamics of melting and solidification, and the latter is due to the layer-wise method of additively made materials. These random defects cause a large fatigue scattering [2,3], where the fatigue life $N$ (i.e., number of loading cycles to failure) varies significantly between identically LPBFed components under same applied stresses [2,4,5]. Notable research on this topic has shown the importance of an optimal process space to avoid formation of porosity [6,7], the relative criticality of defects on the fatigue performance [8,9], and the relationship between the largest expected defect size and the lower bound endurance limit of alloys [10].

The characterization of LPBFed material quality in terms of geometrical defects and fatigue performance is very expensive and time consuming. Fatigue testing coupons require proper heat treatment, machining, and surface finishing prior to fatigue testing [11]. Defect characterization (e.g., morphology and size) can either be done by 2D metallographic inspection or by 3D computed tomography (CT) scans [12], the former of which is labor-intensive and the latter of which is expensive. Optical microscopy and scanning electron microscopy (SEM) for fracture and microstructure inspection are also routine tasks in characterizing LPBFed materials. Fatigue testing time itself contributes largely to experiment cost, all of which makes LPBF fatigue data very expensive and difficult to procure.

---
* Corresponding author, email address: yuebin.guo@rutgers.edu (Y.B. Guo)

Metal fatigue has been historically modeled by empirical expressions and curve fitting [10]. However, such models cannot reliably be applied to LPBFed metal components because of the relatively large uncertainty in their fatigue life due to random geometrical defects, especially in the high cycle fatigue (HCF) region. Almutahhar et al. used the Goodman, Morrow, and Smith-Watson-Topper (SWT) fatigue life prediction models for LPBFed nickel-based alloys data from the literature [13]. Predictions in the HCF region had higher errors than in the low cycle fatigue region, which has also been reported by other LPBF fatigue studies [14–16]. Murakami [17,18] proposed a mechanics-based model for the fatigue strength of metallic materials based on defect size. The lower bound of the fatigue scatter of LPBF materials can be estimated by finding the lowest predicted fatigue strength for a given maximum defect size. This is instrumental for predicting the scatter band of fatigue strength of LPBF materials, which is important for quantifying both the reliability of the material. Kethamukkala et al. [19] combined methodologies from the literature to propose a surface roughness-informed fatigue crack growth (FCG) approach for fatigue life prediction of LPBFed materials. The key aspect of this approach is using an existing surface roughness model to calculate an effective stress concentration factor. The factor is used as an input in a previously proposed stress intensity factor (SIF) formula. A previously proposed time-based (as opposed to cycle-based) FCG model was then integrated to predict fatigue life. The model predictions mostly fell within a factor of 2 of the experimental data, with some scattering left unexplained since random porosity was not investigated.

Machine learning (ML) models have been shown to be useful for predictive tasks when experimental data is massive, or a complex relationship is difficult to establish. ML approaches could also be more efficient than numerical simulations of complicated, nonlinear, coupled physical problems [20,21]. Recently, metal AM fatigue researchers have turned to ML, particularly for fatigue life prediction. Horňas et al. used defect morphology and location data from CT scans with applied stress amplitude as input features to three different ML models for fatigue life prediction of LPBFed Ti6Al4V, finding the best accuracy given by a neural network (NN) model [22]. It is interesting to note that this approach avoids fractography analysis entirely, allowing for nondestructive predictions of fatigue life. Zhang et al. took a different approach by using the LPBF process and post-processing parameters and material tensile property data in a neuro-fuzzy network to predict fatigue life [23]. Initial predictions using the authors' own dataset were within a factor of two of the actual fatigue life, yet validation on unseen data from the literature resulted in poor predictive accuracy. As expected, after incorporating literature data into the training dataset, good predictive accuracy was achieved. This exemplifies the lack of generalizability of data-driven models when data outside the range of the training set is used for testing. Several research groups have used purely data-driven models for FCG predictions. Kamble et al. [24] predicted stage-I and stage-II FCG from experimental data of carbon steel compact tension tests using k-nearest neighbors (KNN), polynomial regression, and ridge regression methods. Mortaza vi and Ince [25] used an NN architecture to predict FCG of short and long cracks of three different alloys from published datasets. Yan et al. [26] used series forecasting and long short-term memory (LSTM) frameworks to forecast brittle crack propagation from phase-field generated randomly distributed cracks. Do et al. [27] applied NN and LSTM models for crack growth forecasting from simulated and experimental data under different material and loading conditions. Hu et al. [28] predicted fatigue life from crack growth of turbine discs, quantifying uncertainty by sampling from uncertainty sources and propagating the uncertainty information to the prediction results.

Physics-informed machine learning (PIML) is an emerging modeling approach. Physical laws or constraints are often incorporated into an NN to guarantee the model predictions comply with physics [29]. It has been shown that PIML models often outperform purely data-driven ML models for engineering applications [30,31]. A few PIML approaches for fatigue performance of AM materials have been reported in the literature [32–38]. Chen and Liu [32] trained an NN model to predict fatigue life distribution of LPBFed Ti-64 for different applied stress amplitudes with available literature data. The resulting plots were referred to as probabilistic *S-N* curves. As a physical constraint, the shape of the predicted *S-N* curves needed to conform to the curvature of experimental *S-N* curves (negative first derivative and positive second derivative). It was shown that without this constraint, the predicted lower-bound fatigue life curve did not have realistic curvature. The log fatigue life was assumed to be normally distributed, and any variability in the retrieved experimental data was neglected. By applying the normal distribution probability functions, the negative log likelihood of the probability of failure was used as the model's loss function. On average, about 80% of the testing data was inside the predicted 95% confidence intervals of fatigue life.

Salvati et al. [33] used a semi-empirical equation based on Murakami and Endo's model [39] for threshold stress intensity factor range based on Vickers hardness $HV$ and defect size $\sqrt{area}$,

$$\Delta K_{th} = 3.3 \times 10^{-3}(HV + 120)\left(\sqrt{area}\right)^{1/3} \qquad (1)$$

as a physical check for their NN model's fatigue life predictions. While residual stress data was not available, applied stress amplitude, as well as defect morphology and location data from CT and fractography analyses were used as model inputs. The training loss was computed as the weighted sum of the NN model loss and the physics loss. The accuracy of the physics-informed NN (PINN) model was compared to that from training solely on the NN model. The PINN model predictions were closer to the actual fatigue life values, with $R^2$ values for the PINN and NN models being 0.591 and 0.322, respectively.

Ciampaglia et al. [34] mimicked Murakami's lower-bound endurance limit model [39],

$$\sigma_w = C_1\,(HV + 120)/\left(\sqrt{area}\right)^{1/6} \qquad (2)$$

in an NN, which gave physically informed predictions of fatigue strength $S$. Basquin's law was used to adapt Equation 2 to incorporate fatigue life as a variable, resulting in

$$S = C_1 C_2 N_f^k\,(HV + 120)/\left(\sqrt{area}\right)^{1/6} \qquad (3)$$

with fatigue life $N_f$, and empirical parameters $C_2$ and $k$. Since Equation 3 shows $S$ proportional to $HV$ and $N_f$, and inversely proportional to $\sqrt{area}$, the PINN model was taught to find microstructural strength $\Phi$, defect size $\Theta$, and fatigue life $N_f^* = C_1 C_2 N_f^k$ parameters that mimic the terms in Equation 3. LPBF process parameter data and fatigue life data were used as model inputs, with the output calculated as

$$S = \frac{\Phi}{\Theta} N_f^* \qquad (4)$$

The predicted fatigue strengths mostly fell within ±150 MPa of experimental values.

In practice, crack size monitoring is critical to ensure safety and avoiding downtime, which is especially important for AM parts having large fatigue scattering. Fatigue life models are popular, but they only give an end result. There is a need for crack size and crack growth prediction methods for LPBFed materials, which have not been adequately addressed in the literature. Konda et al. [40] compiled post-processing, build orientation, and stress intensity factor range $\Delta K$ data of LPBFed Ti6Al4V from the literature for training four different ML models to predict fatigue crack growth rate (FCGR). The extreme gradient boosting algorithm showed the best predictive accuracy of the four models selected. While the results were given in terms of scaled FCGR, no mention of the scaling method was provided to check the validity of the predicted values.

This study aims to develop a PIML model for predicting fatigue crack size and growth rate for LPBFed SS-316L. A unique proof-of-concept approach is presented, using fatigue physics to train a PINN model without the use of experimental FCGR data (i.e., no ground truth). The three novel aspects of this approach are as follows. (1) Physics-informed prediction of fatigue crack size of LPBFed materials has not been investigated; (2) Resonance-based fatigue data has not been studied for its potential as a FCGR process signature; (3) The PIML ability to learn without ground truth data can be leveraged in the AM fatigue field to reduce the need for expensive experimental data. Predicted fatigue crack growth curves are benchmarked with comparable data from the literature to assess the validity of the results. The remainder of this paper is outlined as follows. The procedure for fatigue coupon preparation is explained in Section 2, followed by the fatigue testing procedure in Section 3. Section 4 covers the PIML methodology, and the results are reported and discussed in Section 5. The key findings and an outlook of this research in Section 6 conclude this paper.

## 2. Fatigue coupon manufacturing, preparation, and metrology procedure

Fatigue dog bone coupons were fabricated in three steps. First, a block of SS-316L material was printed using an Aconity machine. Second, a batch of coupons was extracted from the printed block using wire-electrical discharge machining (Wire-EDM). Lastly, the coupons were ground and polished to remove the white layer and heat-affected zone (HAZ) and improve surface roughness of the coupons. Each step is explained in detail as follows.

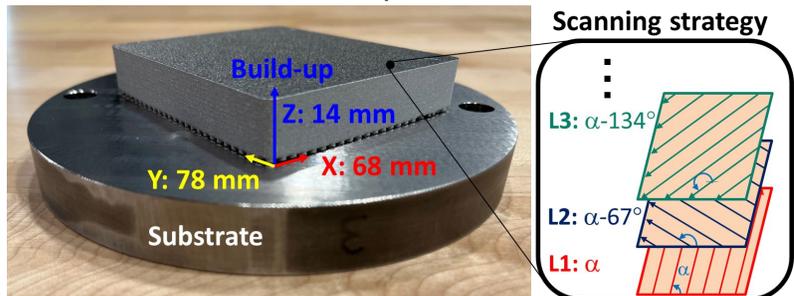

Fig. 1: As-LPBFed SS-316L block on a printing substrate, with dimensions and a schematic of the scanning strategy.

*2.1 Laser powder bed fusion*

Table 1 gives the chemical composition of the SS-316L powder used for the LPBF procedure. The standard in-house process parameters were used (laser power 120 W, laser speed 800 mm/s, hatch spacing 0.08 mm, layer thickness 0.03 mm), giving an energy density of 62.5 J/mm$^3$. A stripe-hatching strategy was employed, using a 1 cm stripe width and a rotation of 67° between layers (see Fig. 1). These parameters were selected through printing optimization to result in minimal block distortion and lack of fusion. Data was taken from coupons from 4 different blocks. All blocks were printed using identical conditions. Little difference was found between the porosity or fatigue performance of these 4 blocks. Thus, it is assumed that all coupons used in this study are comparable. Figure 1 displays an example of an as-LPBFed block.

Table 1: Chemical composition of SS-316L powder.

| C | Cr | Fe | Mn | Mo | Ni | P | Si | S |
|---|---|---|---|---|---|---|---|---|
| 0-0.03% | 16-18% | 61.89-72% | 0-2% | 2-3% | 10-14% | 0-0.045% | 0-1% | 0-0.03% |

*2.2 Wire-electrical discharge machining (Wire-EDM)*

Dog bone coupons were extracted from the LPBFed blocks using wire-EDM. This employs a thin wire electrode (diameter 0.25 mm) to melt away material from an electrically conductive workpiece by high temperature plasma generated from discharges between the electrified wire and the grounded workpiece. The high precision and fine machining capabilities of wire-EDM allow for consistent coupon geometries to be extracted from the as-LPBFed block. This is critical for ensuring each coupon that is fatigue tested is comparable to the entire batch. Machining was conducted in two steps with in-house G-code programs. First, a slicing procedure sectioned the block into a stack of coupons. This step ensures each coupon has the same nominal thickness. Then a profiling procedure created the dog bone profile of each coupon simultaneously with one machining path through the entire stack of coupons. Profiling all coupons simultaneously ensures each dog bone geometry has the same gauge section dimensions.

*2.3 Coupon grinding, polishing, and metrology*

The as-EDMed coupons were ground and polished to remove the EDM-induced white layer and HAZ at the surface layer of the coupons and to reduce the surface roughness prior to fatigue testing. Several stages of sandpaper were used with flushing water to remove most of the roughness, followed by polishing with diamond suspensions. The as-polished surface roughness was checked with a Zeiss LSM 900 laser confocal microscope in the gauge sections (top and bottom) of each coupon. The roughness $R_a <$ 0.2 μm was used as the criterion for sufficient polishing.

Following polishing, the gauge sections were also inspected for any visible porosity using the Zeiss LSM 900. Images of the observed porosity were taken and automatically processed using the ImageJ software to quantify defect size. Size was measured as cross-sectional area of the exposed defect, since only a 2D view of each defect was visible on the coupons. Only porosity above a threshold size of 10 μm$^2$ was recorded, since only the largest defects are of interest in determining fatigue performance and as a practical condition to lessen the time and resources used for metrology.

### 3. Resonance-based fatigue testing: procedure and unique data

Fatigue testing in this study was conducted using a RUMUL Cracktronic resonant testing machine. The operating principle is fully reversed bending of a coupon using an electromagnetically actuated fixture, depicted in Figure 2. The driving mechanism can operate at high frequencies (100's of Hz) for long periods without the need for external cooling by its electromagnetic design, which is advantageous over conventional servo-hydraulic and rotating-bending fatigue systems. Furthermore, the loading frequency is kept at the system resonance frequency by closed-loop control. This has the added benefit of monitoring crack propagation during fatigue testing since the coupon stiffness will change with crack growth, causing a change in the system resonance frequency. This is a unique process signature that has not been explored in the literature for crack growth evaluation. Another advantage of the resonance frequency monitoring over conventional fatigue testing systems is the ability to set the failure criterion as a given drop in frequency for halting fatigue tests instead of testing up to gross fracture. This gives a more practical measure of fatigue performance, since it is desirable to retire any damaged components in use prior to catastrophic failure. A drop of 2 ΔHz is used as the fatigue failure criterion in this study.

The authors hypothesize that the observed drop in resonance frequency during fatigue testing is relatable to crack growth. This relation seems plausible given the similarity in the trends of crack size and frequency drop with time (cycles of loading). Crack growth is known to increase in time following a power law relationship. Experimental data from resonance fatigue testing in Figure 3 shows the same trend for resonance frequency, with a monotonic decrease in frequency, whereas crack size monotonically increases.

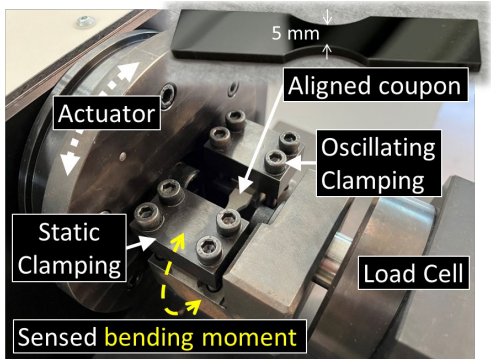
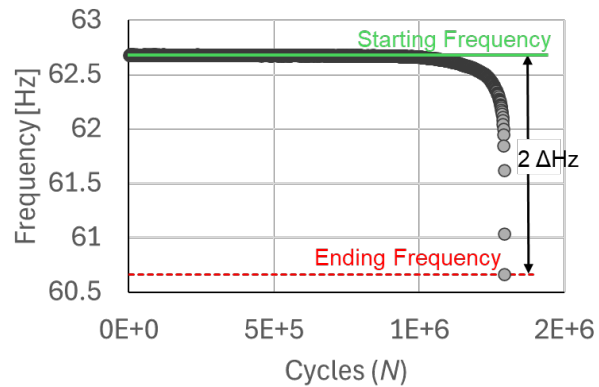

Fig. 2: Resonant fatigue testing fixture with aligned dog bone coupon. An oscillating head moves one side of the fixture. The load cell is built into the static side of the fixture. The coupon is clamped between the fixtures, acting as the system's spring.

Fig. 3: Monotonically decreasing resonance frequency during a fully reversed fatigue test of LPBFed SS-316L. Failure criterion is a drop in resonance frequency of 2 ΔHz.

A flowchart of the fatigue testing procedure is given in Figure 4. A dog bone coupon is aligned in the fixture using custom-made shims. A torque wrench is used to secure the coupon with balanced, consistent pressure on all clamping bolts. The desired bending moment (corresponding to a desired stress amplitude $\sigma_a$) is input to the software of the machine's control computer. After starting the test, the computer will receive loading and frequency signals from the machine to monitor the resonance frequency and the applied loading. The data is recorded every 1000 cycles. As the test progresses, any change in the resonance frequency from the initial reference value is recorded. After the resonance frequency deviates from the initial reference value by 2 Hz, the test is automatically ended. If this does not occur after $10^7$ cycles, the test is halted as a runout.

After fatigue failure occurs, the coupon is fractured but still whole. To inspect the fracture surface for the fracture source (i.e., a geometrical defect), the coupon is manually fractured into two parts about the existing fatigue crack. Then optical and scanning electron microscopy are employed to observe and measure the defect size.

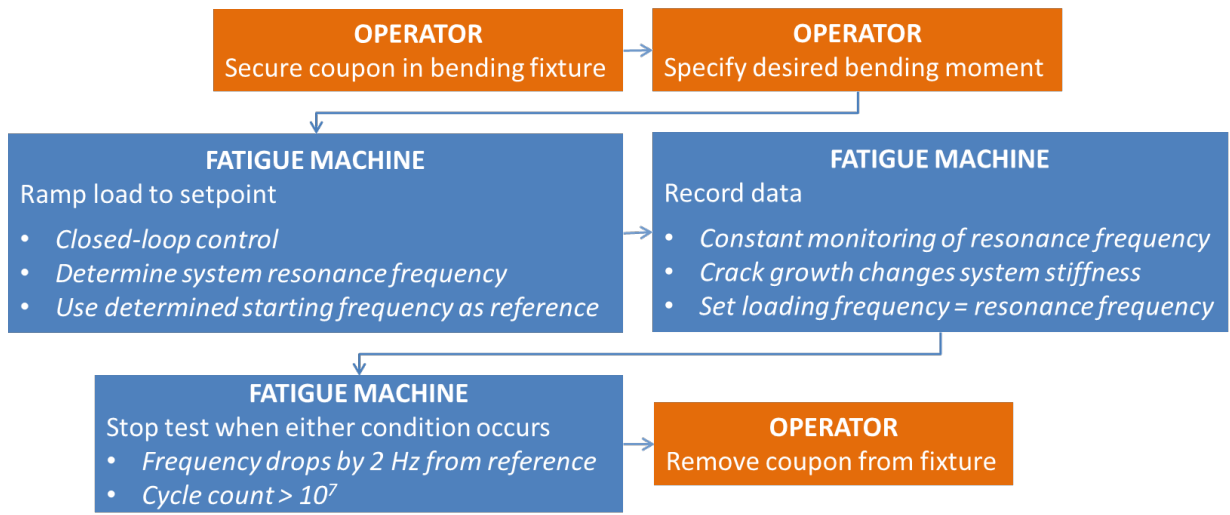

Fig. 4: Flowchart of fatigue testing procedure.

## 4. Physics-informed machine learning (PIML) of crack size

Regular inspection is typically scheduled to monitor and find any fatigue cracks in the lifecycle of a component. An improvement would be to track any crack growth in real time via appropriate signals and implement autonomous inspection. This would reduce the risk of catastrophic failures and the man-hours used for inspection.

The authors hypothesize that the observed drop in resonance frequency during fatigue testing is a process signature to crack growth. No known fundamental law exists for this relationship. Moreover, no experimental crack size data is available for the fatigue tests conducted since no existing measurement method (e.g., optical) is compatible with the resonant fatigue testing setup. To discover this possible relationship, PIML can be explored. By using the resonance frequency data from the fatigue tests in conjunction with a physical law for crack growth, a PIML model can be trained to predict fatigue crack size.

In general, PIML models learn from the integrated governing equations and experimental data by computing error between the predicted and experimental values (data loss) and computing residual error in the governing equations (physics loss). In this study, since no experimental crack size data is given, only physics loss will be used to train the model. The next sections discuss the physical law used to train the PIML model, the model details, and the data preparation procedure.

*4.1 Paris's law*

Paris's law is an empirical law governing fatigue crack growth during steady crack propagation. Usually, Paris's law is used for fatigue life calculations by integrating over an initial crack size to a given crack size (e.g., largest acceptable crack size for safe design). In this study, it is of interest to find crack size during fatigue testing, a feat which is difficult to achieve in practice. The rate of crack growth is related to the stress intensity factor range as

$$da/dN = C(\Delta K)^m \tag{5}$$

where $a$ is the crack size (in m, or more typically mm), $N$ is the number of cycles of loading, $\Delta K$ is the stress intensity factor range (in MPa·m$^{1/2}$), and $C$ and $m$ are fitting constants. The crack growth coefficient $C$ typically is reported with units $\frac{\text{mm/cycle}}{(\text{MPa}\sqrt{\text{m}})^m}$ to balance the units of Equation 5.

The parameters $a$, $\Delta K$, $C$, and $m$, are the unknowns. Experimental data for $N$ is available. With the crack size $a$ being primarily of interest, three values ($\Delta K$, $C$, and $m$) need to be predicted by the PIML model to make computations for $a$ with Equation 5. Care must be taken with how these values are scaled as outputs due to the magnitude and unit scaling that applies to each. Numerical instability can occur in ML models when the model parameters have magnitudes larger than unity. Values of $C$ can easily reach the nanoscale, if not smaller, and $m$ typically has values larger than unity. $\Delta K$ also has a unique scale as well as units to consider. The next section will discuss how these parameters are scaled in the PIML procedure.

*4.2 Nondimensional analysis of Paris's law*

When using a governing equation for evaluating physics loss in an ML model, it is helpful to perform normalization. This allows for unitless values with ubiquitous magnitude scaling to be predicted by the model. To nondimensionalize Equation 5, each variable must be scaled by an appropriate scaling factor of like units. The variables are the crack size $a$ (m), number of cycles $N$, and the stress intensity factor range $\Delta K$ (MPa·m$^{1/2}$). Appropriate scaling factors are selected as follows.

- The crack size can only grow as large as the coupon gauge width $W = 5$ mm. The normalized crack size is thus

$$a^* = a/W \tag{6}$$

- The number of cycles to failure has the practical limit set by the runout condition $N_r = 10^7$ cycles. The resulting normalized cycle value is

$$N^* = N/N_r \tag{7}$$

It should be noted that normalizing $N$ is not necessary in this study, since this information is known and not something predicted by the PIML model. However, it is still mentioned here for completeness.

- Stress intensity factor range is dependent on stress amplitude $\sigma_a$ (MPa) and $a$ by

$$\Delta K = 2Y\sigma_a\sqrt{\pi a} \tag{8}$$

where $Y$ is a unitless geometry factor. The same scaling factor for $a$ applies here, and $\sigma_a$ can be normalized by the lower bound endurance limit $\sigma_w$ given by Murakami's fatigue strength model (see Equation 2). $\sigma_w$ is calculated from the material hardness $HV$ and the maximum defect size $\sqrt{area}$. In this study, $\sqrt{area}$ is the size of the defect at the fracture surface. This gives the normalized stress intensity factor range

$$\Delta K^* = \Delta K / (\sigma_w \sqrt{W}) \tag{9}$$

Here, the stress intensity factor is unknown and will be an output (prediction) of the PIML model, along with Paris's law constants $C$ and $m$. Outputs from the model will be in normalized form (e.g., $\Delta K^*$). In the case of $C$ and $m$, the values that are yielded from the model do not have any nondimensional form but rather require magnitude scaling. These are then scaled back to the correct dimensions (e.g., with Equation 9) for evaluating the physics loss using Paris's law (Equation 5). The next section discusses the details of the data used by the PINN model for predicting crack size.

*4.3 PINN input data description and preparation*

Before introducing the PINN model architecture, it is necessary to curate data first. Four input variables are used by the PINN model for predicting crack size. These are prepared from the following experimental data: defect size $\sqrt{area}$ (µm), stress amplitude $\sigma_a$ (MPa), cycles of loading $N$, and resonance frequency $f$ (Hz). Each coupon has one (scalar) value for $\sqrt{area}$ and $\sigma_a$ since each coupon has one measured defect size at the fracture surface and one load level during fatigue testing. Each coupon has a vector of frequency data and a vector of cycle data (e.g., see Figure 3). These are time series data $N_i$ and $f_i$ for $i = 1, 2, \ldots, n$, where $n$ is the final index of data collected for the duration of each coupon's fatigue test. The first three data are scaled logarithmically (i.e., $\log_{10}\sqrt{area}$, $\log_{10}\sigma_a$, $\log_{10}N_i$) to scale the values to be within the same order of magnitude. One further step is needed to scale the cycle data, namely, min-max normalization. This is needed to ensure the model learns the dependence of crack size on cycle count. This is because defect size and stress amplitude are constant for each fatigue test, while cycle count is variable. This means that the change in cycles should be of significant magnitude; otherwise, small changes will be overlooked by the model and will not have any effect on the predicted crack size. As for the resonance frequency data, several transformation steps are required, which are explained as follows.

The raw data (see Figure 3) has several characteristics that would otherwise hinder the PINN model from learning if not transformed. Firstly, the order of magnitude of the values should be reduced to match that of the other data. Secondly, a monotonically increasing trend is needed to better indicate the physical process of crack growth, which requires the decreasing frequency trend to be flipped. Thirdly, the majority of the data collected is not indicative of the stable crack growth region modeled by Paris's law. Only the data that shows the decline in frequency should be retained, while all prior data from the crack initiation stage of the fatigue test (the linear region of the frequency curve in Figure 3) should be excluded. The steps for performing these transformations, which are slightly adapted from the authors' previous work [4], are given as follows.
1. The data is smoothed by taking a moving average.
2. The transition point between the linear and nonlinear parts of the frequency data is determined by iteratively fitting a linear line to increasingly larger portions of the data.
    a. Linear regression is applied to an initial number of data points $\omega$.
    b. The residual sum of squares $RSS$ for linear regression is calculated as
    $$RSS_\omega = \sum_{i=1}^{\omega} (f_i - \hat{f}_i)^2 \qquad (10)$$
    where $\hat{f}_i$ is the regressed frequency value.
    c. The initial $RSS$ value is saved as the current minimum $RSS_{min}$.
    d. Consecutively add data points (i.e., increase $\omega$) to update the regression line and re-solve $RSS$.
    e. If $RSS_\omega < RSS_{min}$, update the saved minimum value of $RSS$ by the current value.
    f. For a given factor $\alpha$ of the fit error, a threshold error level can be set, above which the transition from linear to nonlinear regime is indicated. I.e., if $RSS_\omega \geq \alpha \cdot RSS_{min}$ then stop associating data with the linear region of the data. Use the previous value of $\omega$ as the index at which to cut off the linear portion of the frequency data from the nonlinear portion.
3. Discard $f_i$ for $i \leq \omega$.
4. The remaining data is transformed into a new variable $\Delta f_i$ by the following transformation
$$\Delta f_i = \frac{-(f_i - f_{max}) - f_{min}}{f_{max} - f_{min}} \qquad (11)$$
This first finds the magnitude of the drop in frequency (by the $-(f_i - f_{max})$ term) and then applies min-max normalization to scale the values in the range $[0,1]$.

Figure 5 gives an example illustration of the transformation of the resonance frequency data. 10 datasets were available, all of which were used individually as 10 different trials for model development. The next section outlines the PIML model details.

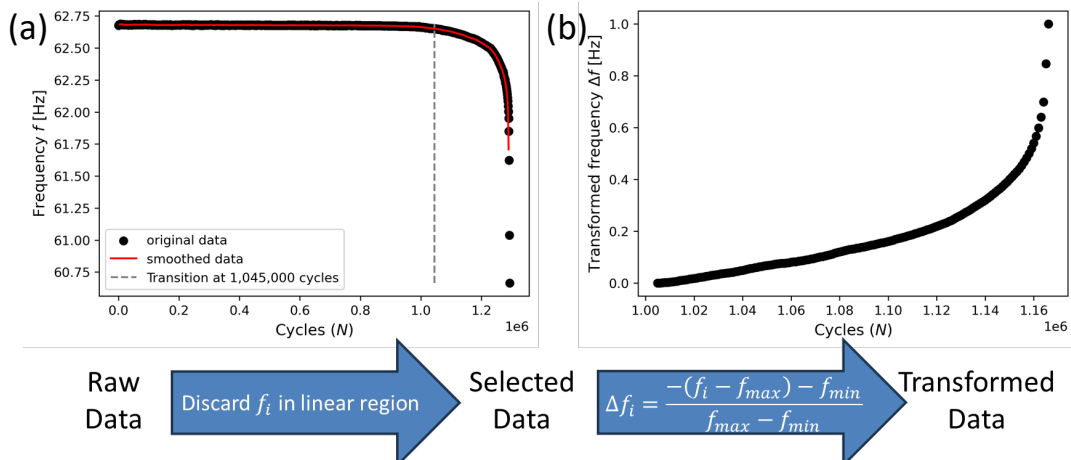

Fig. 5: Procedure for resonance frequency data preparation. (a) Selection of nonlinear data (steps 1-3 of section 4.3). (b) Transformation of data to normalized frequency drop $\Delta f$ (step 4 of section 4.3).

*4.4 PINN model architecture and loss function*

A PINN model was created using the PyTorch Python package. Figure 6 gives a visual overview of the PINN model characteristics. A feed-forward neural network (NN) with four inputs and four outputs is constructed with two hidden layers. Sigmoid activation $\sigma(\cdot)$ is applied to the outputs to keep all predicted values positive. The mathematical background of NN operations is not discussed here since NN models are very commonplace ML tools. It is more appropriate to focus on the novel physics-informed approach of this study, which uses Paris's law in the model loss function. The (total) loss function is defined as

$$Loss = w_1 \cdot L_{IC} + w_2 \cdot L_{BC} + w_3 \cdot L_{mon,a} + w_4 \cdot L_{mon,K} + w_5 \cdot L_{RSS} \tag{12}$$

which has five terms, each with weight $w_i$. Each loss term is explained as follows.

$$L_{IC} = |\Delta K_{i=1} - \Delta K_{th}| \tag{13}$$

The initial condition loss $L_{IC}$ is the absolute difference between the empirical threshold stress intensity factor range $\Delta K_{th}$ (given by Equation 1) and the stress intensity factor range predicted from the first set of input data (i.e., chronologically first with respect to the time series data of frequency and cycles). This term is necessary to ensure the predicted crack growth curve starts from a sensible magnitude. A good starting point for $\Delta K$ should be slightly larger than the threshold value. Thus, this difference should be minimized by the model.

$$L_{BC} = \left|\log_{10} W - \log_{10}\left(\sum_{i=1}^{n}[C(\Delta K)^m]_i \cdot \Delta N_i\right)\right| \tag{14}$$

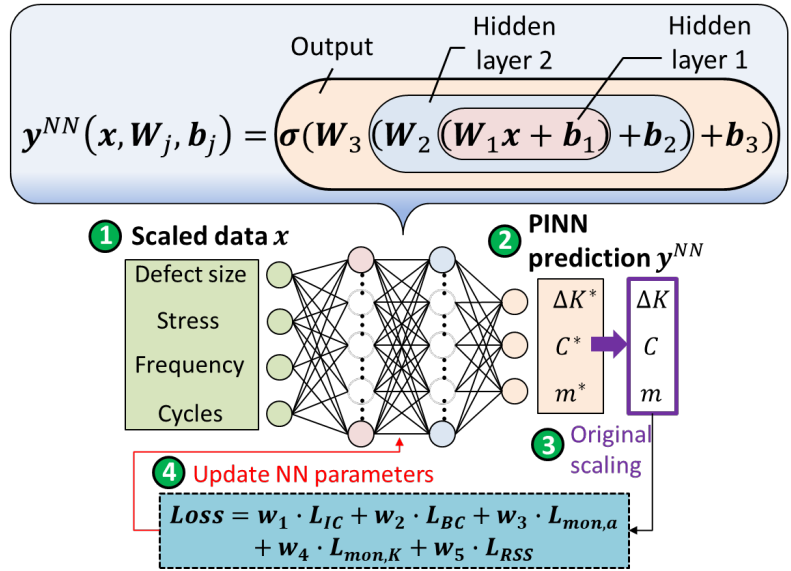

Fig. 6: Overview of PINN model architecture and loss function. 1) The input data is scaled and given to the NN. 2) The predicted outputs, after applying the Sigmoid activation $\sigma(\cdot)$, are the nondimensional stress intensity factor range $\Delta K^*$ and the scaled values of the constants $C$ and $m$ (i.e., $C^*$ and $m^*$). 3) The parameters are transformed back to the original scaling. 4) Paris's law is used to evaluate the compatibility of the prediction with crack growth physics, with the loss being used to update the NN weights $W_j$ and biases $b_j$.

The boundary condition loss $L_{BC}$ is the absolute difference between the log-scaled gauge width $W$ and the log-scaled final crack size. The final crack size is found by discrete integration of the predicted right-hand side of Paris's law with respect to $N$. Log-scaling ensures the magnitude of this difference is sufficiently large to result in efficient updates of the model parameters. Otherwise, the relatively small scale of the crack size ($10^{-3}$ m) results in an insignificant error term magnitude, which would have to be magnified by the weight $w_2$. The combination of scaling and applied weight worked well to promote learning.

$$L_{mon,a} = \sum_{i=1}^{n-1} (da)_i \cdot \begin{cases} -1, & da < 0 \\ 0, & da \geq 0 \end{cases} \quad (15)$$

$$(da)_i = [C(\Delta K)^m]_{i+1} - [C(\Delta K)^m]_i \quad (16)$$

The crack size monotonicity loss $L_{mon,a}$ computes the error in the crack growth rate due to negative $da/dN$ values. Physically, crack size should monotonically increase. Thus, the change in crack size between consecutive PIML predictions should not be negative. All negative slope values $da < 0$ are summed. The sign is flipped for a positive loss value. It is possible when the model parameters update that this loss term disappears, i.e., $L_{mon,a} = 0$. When this happens, the model has learned the underlying physics. Therefore, even if this loss term is only present initially during model training, its disappearance does not indicate it was not necessary. Furthermore, since multiple loss terms are used, it is possible for the model to update parameters to satisfy the other terms and by chance result in a return of the monotonicity loss. This loss term is always computed during training to ensure any error is corrected when it arises.

$$L_{mon,K} = \sum_{i=1}^{n-1} (dK)_i \cdot \begin{cases} -1, & dK < 0 \\ 0, & dK \geq 0 \end{cases} \quad (17)$$

$$(dK)_i = [\Delta K]_{i+1} - [\Delta K]_i \quad (18)$$

The stress intensity factor range monotonicity loss $L_{mon,K}$ is identical in form and function to $L_{mon,a}$. Physically, $\Delta K$ monotonically increases with crack size. Therefore, a monotonicity error term is included to ensure the model learns this.

$$L_{RSS} = RSS\left(\log_{10} \Delta K, \log_{10}[C(\Delta K)^m]\right) \quad (19)$$

The residual sum of squares loss $L_{RSS}$ finds the error in the linear regression of the predicted $\Delta K$ and the predicted crack growth rate $da/dN$ from the right-hand side of Equation 5. At log scale, Paris's law gives a linear relationship between $\Delta K$ and $da/dN$. Thus, this loss term trains the model to conform to this physical law by giving the error in residual of Paris's law for the predicted crack growth curve.

Each loss term is necessary for guiding the model towards a physically correct correlation between the input data and the desired outputs. It is important to note that there is no ground truth data for comparison with the predicted values (i.e., no targets/labels). As this is a proof-of-concept study, the authors understand the importance of future experimental validation for a direct measurement of accuracy. As noted earlier, no method of collecting crack growth data is yet available for the current fatigue testing procedure. This challenge is left for future consideration.

## 5. Results and Discussion

The PINN-predicted crack growth curves for dataset 1 are displayed in Figure 7. The model learned from the physical loss terms only since there was no target data for comparison. Because of this, the data could be used as many times as necessary for training the model since the data itself was not being learned. 50 epochs were sufficient for the total loss to converge. As is seen in Figure 7a, a near-perfect linear trend in the log-log scaled Paris's law curve. This is a direct indication of the model's ability to capture the physics by learning from the loss terms, specifically the one in the $L_{RSS}$ loss. It is worth noting that experimental Paris's law graphs do not show as high linearity in the data since some random scattering and deviations from the linear trend occur. This random scattering is beyond the scope of the present work. Further improvement to this work could seek to incorporate this inherent variation as a learnable parameter. It would be interesting, for example, to relate variation in the crack growth behavior to the LPBF process to seek a better understanding of how the manufacturing process influences the resulting fatigue behavior. Figure 7b shows the computed crack size from the predicted crack growth rate, with the final crack size overshooting the boundary condition of coupon gauge width. Since the actual crack size is unknown, the gauge width is assumed to be a physical limit to the crack growth size across the coupon surface. However, the predicted result is not far from this assumed limit and only exceeds the expected final crack size in the last two data points. This is due to the rapid transition from stable crack growth to rapid fracture, which is an important phenomenon captured by the model. Thus, even with some error in the final crack size, this result is consistent with fatigue physics.

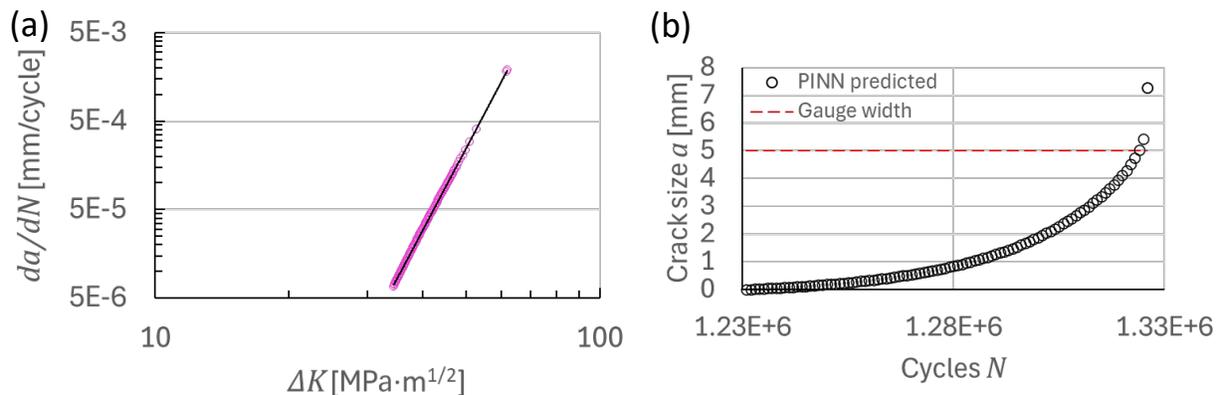

Fig. 7: PINN-predicted crack growth curves for dataset 1. (a) Crack growth rate computed from the predicted stress intensity factor range and Paris's law parameters. (b) Computed crack size from integration of the crack growth rate. Fatigue coupon gauge width of 5 mm shown for comparison with PINN prediction of final crack size.

To better understand the impact of each loss term during model training, Figure 8 shows the loss curves of each individual loss term as well as the total loss over the 50 training epochs. Both linear scaling and log scaling of the losses are given for better visualization of the evolution and contribution of each term during the training process. As mentioned in Section 4.4, the monotonicity loss terms $L_{mon,a}$ and $L_{mon,K}$ only appear when the monotonicity condition for crack size and stress intensity factor range is violated. As such, only the first two epochs show nonzero loss values for $L_{mon,a}$ and only the second epoch resulted in a nonzero loss for $L_{mon,K}$. The fact that all other epochs had zero loss for these terms shows the model learned the monotonicity condition early in the training process. The other three loss terms $L_{IC}$, $L_{BC}$, and $L_{RSS}$ converge to stable values after about 40 epochs. The initial condition loss $L_{IC}$ showed the most gradual and most stable decline of all the loss terms, while both the boundary condition loss $L_{BC}$ and the residual sum of squares loss $L_{RSS}$ showed more fluctuation while reaching convergence. This is particularly noticeable in the last 10 epochs for $L_{BC}$ where the loss oscillates about a constant level averaging at about 0.36. Adapting the learning rate hyperparameter could improve stability in the loss convergence but was not necessary to achieve the present results.

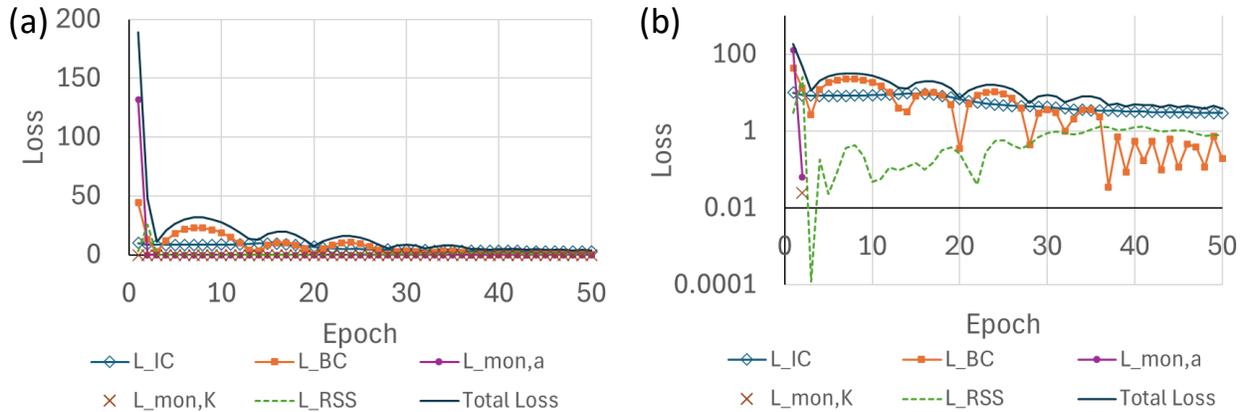

Fig. 8: Contributions of each loss term to the total training loss. (a) Linear scaling of loss values. (b) Log scaling of loss values.

Similar results are shown for the other nine datasets in Figure 9a, where all 10 PINN predicted Paris's law curves are plotted together for comparison. All fatigue coupons were manufactured from the same material with the same process parameters. In a typical fatigue crack growth test with compact tension specimens, there should be little difference between the crack growth curves of a given material. Figures 9a and 9b show a dependence on dataset size, which is shown to be related to the stress amplitude $\sigma_a$ of each dataset. Larger stress amplitudes typically lead to earlier failures, resulting in faster crack growth and therefore less data. Physically, this results in larger crack growth rates, which are captured by the PINN model.

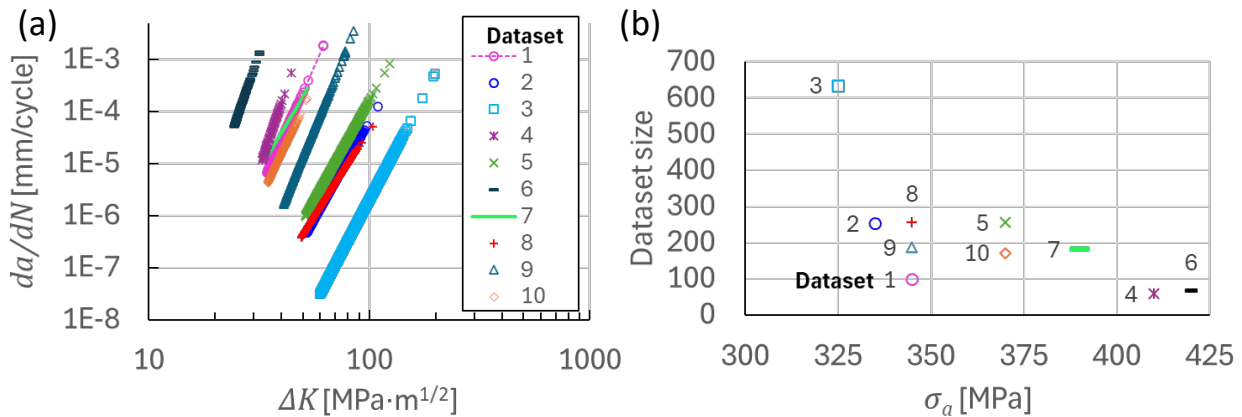

Fig. 9: (a) Paris's law curves from individual model trainings using 10 separate input datasets. Range of crack growth shows dependence on dataset size. (b) Dependence of dataset size on stress amplitude $\sigma_a$, with dataset identifier labels next to each data point.

Each input dataset has constant stress amplitude and defect size inputs. Thus, the optimized NN parameters were specifically tuned for these inputs based on the computed physical losses, resulting in different crack growth curves for each input dataset. Combining the datasets should therefore result in more uniform results since having a variety of values for each input feature would result in different physical losses and drive the model to update its parameters for more input cases. This is indeed shown to be true in Figure 10a, where 10 Paris's law curves are predicted, one for each input dataset, after combining all datasets to train a single model. Each dataset was given as input data for 5 epochs of training for a total of 50 epochs of combined training. All curves overlap and show high similarity, regardless of dataset size. Comparison of the initial condition of starting stress intensity factor range for the individual vs. combined training methods shows the consistency of the combined training method results (see Figure 10b). The relative difference between the empirical threshold values and the PINN predicted initial values is relatively constant. Small variations in the starting value are expected since the measured maximum defect sizes and material hardness values of each fatigue testing coupon are similar in magnitude.

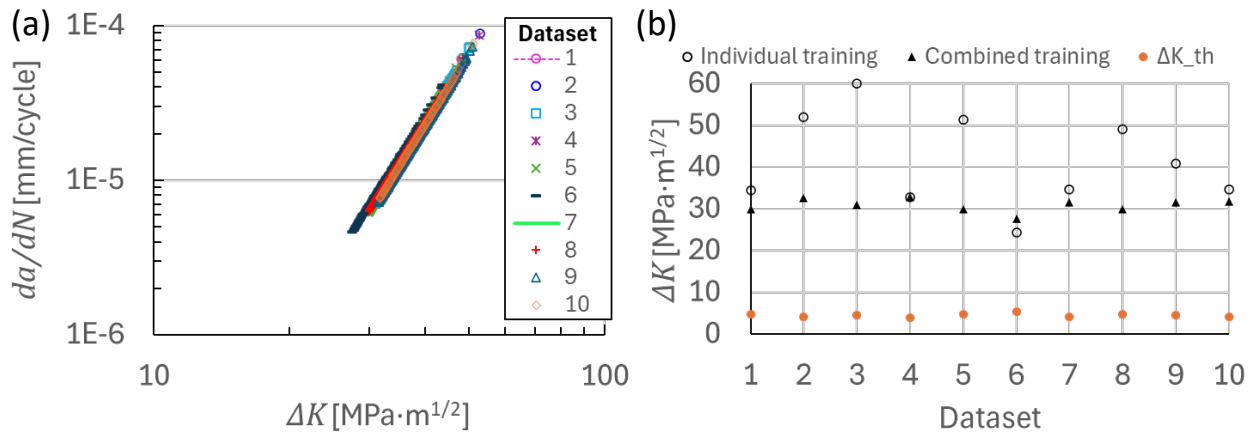

Fig. 10: (a) Paris's law curves from combined training of a single model using all 10 input datasets. (b) Comparison of starting stress intensity factor range values with empirical threshold stress intensity factor range given by Equation 1.

Several other results can be compared between the individual and combined training methods. Figure 11 compares the final crack sizes and Paris's law constants $C$ and $m$, found by linear regression, for each crack growth curve of the individual and combined training methods. Paris's law parameter values for LPBFed SS-316L from the literature are given in Table 2. Of the few studies that report these values, the manufacturing and fatigue procedure in this study aligns best with that of Kluczyński et al. [41], with similar LPBF energy density values (~7% difference) and the same procedure of parallel crack growth direction to the build-up direction. The results from Kluczyński et al. are therefore plotted along with the PINN predicted values from this study in Figure 11 for a fair comparison. The combined training results give very similar values of Paris's law constants, whereas the individual training results show large variations. The combined training method gives final crack sizes closer to the boundary condition than the individual training method for 6 of the 10 input datasets, with one additional dataset (dataset 10) giving very similar results for both methods (see Figure 11a).

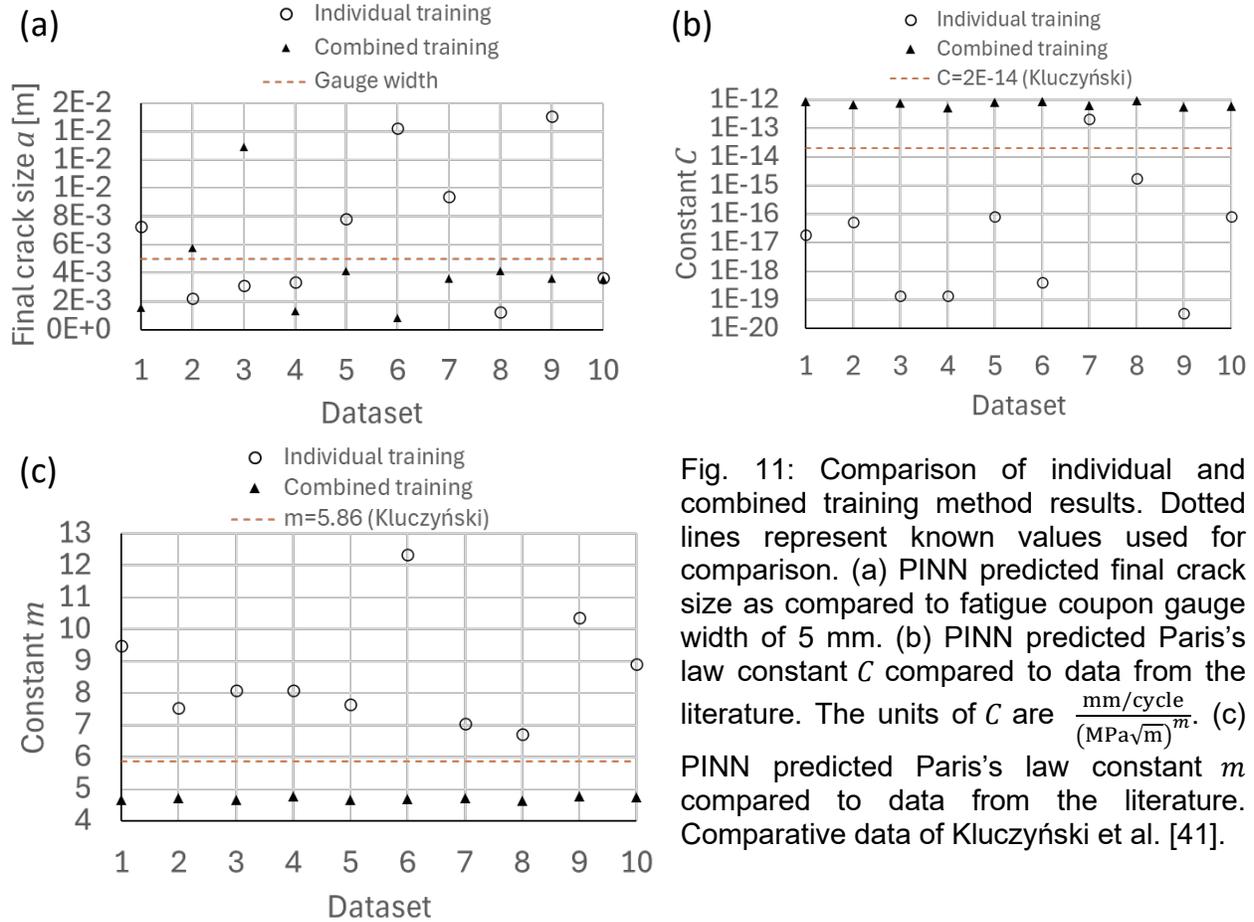

Fig. 11: Comparison of individual and combined training method results. Dotted lines represent known values used for comparison. (a) PINN predicted final crack size as compared to fatigue coupon gauge width of 5 mm. (b) PINN predicted Paris's law constant $C$ compared to data from the literature. The units of $C$ are $\frac{\text{mm/cycle}}{(\text{MPa}\sqrt{\text{m}})^m}$. (c) PINN predicted Paris's law constant $m$ compared to data from the literature. Comparative data of Kluczyński et al. [41].

Table 2: Published experimental data of Paris's law constants for LPBFed SS-316L. Units of $C$ are $\frac{\text{mm/cycle}}{(\text{MPa}\sqrt{\text{m}})^m}$.

| Energy Density [J/mm³] | Crack growth direction vs. build direction | $C$ | $m$ | Ref. |
|---|---|---|---|---|
| 58.64 | parallel | 2.00E-14 | 5.86 | [41] |
| -- | parallel | 2.12E-09 | 3.37 | [42] |
| 95.2 | perpendicular | 8.63E-10 | 3.46 | [43] |
| 95.2 | perpendicular | 1.74E-10 | 3.55 | [43] |
| 95.2 | perpendicular | 2.50E-10 | 3.76 | [43] |

The combined training results show close agreement with the Paris's law constants from the literature. It is known that the crack growth rate of a material is dependent on the stress ratio $R$ due to crack closure [44]. As $R$ decreases, the fatigue crack growth curve shifts rightward, resulting in larger values of $C$ over the case of repeated loading ($R = 0$). The PINN predicted value of the Paris's law coefficient $C$ for the combined training case agrees with this physical phenomenon. The value of $C$ from the study by Kluczyński et al. resulted from experiments with $R = 0.1$, while the experiments in this work were conducted with $R = -1$. Accordingly, the PINN predicted value of $C$ is larger than the value from literature, and thus the stress ratio effect is observed. As for the Paris's law power $m$, the PINN predicted values for the combined training case and the experimental value from the literature are in close agreement (predicted average of 4.70 vs. experimental value of 5.86).

## 6. Conclusions

A novel nondimensionalized PIML model has been developed to predict fatigue crack growth with zero ground truth (i.e., no targets/labels used for model development). The unique application of resonance frequency as a fatigue process signature with Paris's law shows potential as an alternative method of determining crack size in fatiguing components. This can especially help the metal AM industry, where random porosity causes large fatigue scattering, making it insufficient and risky to solely rely on predictive fatigue life models. The important findings are as follows.

- Resonance frequency during fatigue testing can be used with physics to relate to fatigue crack growth.
- A novel nondimensionalized PIML framework is proposed using Paris's law without experimental data of fatigue crack growth
- Paris's law constants can be learned without any prior knowledge of material properties aside from material hardness.
- Predicted crack growth curves match the physical trend and give the correct range of crack size.
- Predicted Paris's law constants are consistent with comparable experimental data from the literature.
- Datasets of input data can either be used individually to develop case-specific models or combined to train a single model for material-specific crack growth prediction.

Several future directions would be pursued. First, experimental crack growth data may further validate the predicted crack growth curves. This would involve proposing a methodology for using the existing fatigue testing setup for crack growth measurements. It is noted that such a task is difficult to accomplish for such relatively small fatigue coupons, especially since the existing setup is not made for crack growth observation. Second, expanding the available data to include different AM materials, specifically adding microstructure or material property data, could also be explored for the purpose of training the PIML model to learn different materials. This could be an opportunity to investigate transfer learning of the model parameters by training on datasets of different materials. Additionally, further experimental data can be curated for alternative stress ratios to have more generalizable predictions for different fatigue loading cases. Including random scattering as a learnable parameter could also be considered in a future work. For example, relating variation in the crack growth behavior to the LPBF process could shed light on the relationship between the manufacturing process and the fatigue behavior.

## Acknowledgements

The authors would like to thank the financial support of the National Science Foundation under the grants CMMI-2412395.